%&amstex
\documentstyle{faa_eng}
\pageno=6

\hoffset=15pt
\voffset=15pt

\define\hhL{\rlap{\hskip1.2pt\lower-10.5pt\hbox{\special{em:graph
hat_1.pcx}}}\Cal L}
\define\Lhh{\rlap{\hskip1.2pt\lower-10.5pt\hbox{\special{em:graph
hat_2.pcx}}}\Cal L}
\define\Lh{\rlap{\hskip1.2pt\lower-9.5pt\hbox{\special{em:graph
hat_3.pcx}}}\Cal L}
\define\ovL{\ovl11{\Cal L}}
\define\ovU{\ovl11U}

\title{29}{1}{95}{On Whitham's Averaging Method}

\authorinfo{A.~Ya.~Maltsev and M.~V.~Pavlov\ft{Partially supported by
the Russian Foundation for Basic Research under grants Nos.
94-01-01478 and 93-011-168.}} {517.944}{Landau Institute of
Theoretical Physics, Russian Academy of
Sciences}{29}{1}{7--24}{February 20, 1994}

%\pred{V. M. Volosov}

\bh Introduction\endbh Whitham's averaged equations~\cite1 for a
nonlinear evolution system describe slow modulations of parameters on
a family of periodic traveling wave solutions (or on families of
multiphase quasiperiodic solutions, which are so far knows to exist
only for integrable equations) and are a system of hydrodynamic type
\cite{2, 3}, that is, of the form
$$
U_T^i=V_j^i(\ovU\)U_X^j,\qquad i,j=1,\dots,N\tag1
$$
(we consider only the spatially one-dimensional case), where
$\ovU=(U^1,\dots,U^N)$.

The original evolution system is usually Lagrangian or Hamiltonian,
and this property is then inherited by the equations of slow
modulations. A Hamiltonian theory of systems~(1) was constructed by
B.~A.~Dubrovin and S.~P.~Novikov~\cite2 and, then, successfully used
by S.~P.~Tsarev~\cite7 to integrate Hamiltonian systems reducible to
the diagonal form
$$
U_T^i=V^i(\ovU\)U_X^i,\qquad i=1,\dots,N\.\tag2
$$

In the present paper we give an exposition and justification of some
topics in Whitham's averaging theory. Namely, we consider classical
single-phase Whitham averaging for a Lagrangian system in which some
of the fields (denoted by $u^i$) have a direct physical meaning,
whereas for the other fields (denoted by $\ph^\a$) only the
$x$-derivatives $\ph_x^\a(x)$ are physically meaningful. The
averaging is carried out on a family of periodic traveling waves of
the system rewritten in the coordinates $u^i(x)$,
$q^\a(x)=\ph_x^\a(x)$. To average such systems directly on the basis
of the Lagrangian formalism, Whitham~\cite1 proposed the pseudophase
method, which permits one to obtain the averaged equations in the
Lagrangian form and then to construct the corresponding Hamiltonian
structure \cite6 (for the case in which pseudophases are lacking, the
Hamiltonian formalism for the Whitham equations was constructed by
Hayes~\cite{11}). However, there is an alternative approach. It is
based on the fact that in the variables $(u^i(x),q^\a(x))$ the system
is Hamiltonian and possesses conservation laws of the form
$S_t^i=R_x^i$, which correspond to the energy and momentum
conservation laws and to the annihilators of the Poisson bracket. By
using these conservation laws, we can rewrite the equations of slow
modulations in the form
$$
\lan S^i\ran_T=\lan R^i\ran_X,\tag3
$$
where $\lan\,\dots\,\ran$ stands for the averaging on the family of
traveling waves \cite1. The Hamiltonian structure of Eqs.~(3) can be
obtained by averaging the original Hamiltonian structure by the
Dubrovin--Novikov method \cite{2, 3}, but the general proof of the
Jacobi identity for the averaged Poisson bracket is lacking \cite4.
In the present paper we prove that 1) both methods lead to the same
equations of slow modulations (this was proved in \cite4 by the WKB
method) and 2) the Poisson bracket averaged by the Dubrovin--Novikov
method coincides with that obtained from the averaged Lagrangian
formalism (thus, the Jacobi identity for the Dubrovin--Novikov
bracket is proved in this case). Furthermore, we prove that if the
system has additional conservation laws of the form $S_t=R_x$, then
this averaging results in equations consistent with~(3).

\bh1. Averaging of conservation laws and the pseudophase method\endbh
Let $a^q(x,t)$ be the field functions, and let the action
$$
S=\int L[a^q(x,t)]\,dt=\iint{\Cal L}(a^q,a_t^q,a_x^q,a_{tt}^q,
a_{xx}^q,a_{xt}^q,\dots)\,dx\,dt,\tag4
$$
$q=1,\dots,N$, be given.

Suppose that an action $\wh S\in T^m\:(a^q)\to(\acute a^q)$ of an
$m$-dimensional Abelian group $T^m$ (not necessarily compact) on the
manifold of fields is given, the orbits of this action are
$m$-dimensional, and the Lagrangian is invariant, that is,
$$
\Cal L(a^q,a_t^q,a_x^q,\dots)=\Cal L(\acute a^q,
\acute a_t^q,\acute a_x^q,\dots)\.\tag5
$$

By Noether's theorem, each symmetry of the Lagrangian (including
translational invariance and independence of time) yields a
conservation law. It is easy to verify that these laws have the form
$$
\bigg[\sum_{n\ge1}n\sum_qa_{nt}^q\,\frac{\pa\Cal L}{\pa a_{nt}^q}
-\Cal L\bigg]_t+\bigg[\sum_{n\ge1}n\sum_q
a_{(n-1)x,t}^q\,\frac{\pa\Cal L}{\pa a_{nx}^q}\bigg]_x
+[\,\dots\,]_{tt}+[\,\dots\,]_{xx}+[\,\dots\,]_{xt}=0\tag6
$$
(conservation of energy) and
$$
\bigg[\sum_{n\ge1}n\sum_q a_{(n-1)t,x}^q\,\frac{\pa\Cal L}{\pa a_{nt}^q}
\bigg]_t+\bigg[\sum_{n\ge1}n\sum_q a_{nx}^q
\frac{\pa\Cal L}{\pa a_{nx}^q}-\Cal L\bigg]_x
+[\,\dots\,]_{tt}+[\,\dots\,]_{xx}+[\,\dots\,]_{xt}=0\tag7
$$
(conservation of momentum); here $a_{nt}^q\equiv\pa^na^q\!/\pa t^n$,
$a_{nx}^q\equiv\pa^na^q\!/\pa x^n$.

Before writing out the other conservation laws, let us pass to new
field variables, $(a^q)\to(u^i,\ph^\a)$, where $u^i$ are constant on
the orbits of $T^m$ and $\ph^\a$ are parameters on these orbits (and
coincide with the parameters on the group itself).

In these variables the other conservation laws acquire the form
$$
\bigg[{\pa{\Cal L}\over\pa\ph_t^\a}\bigg]_t+
\bigg[{\pa{\Cal L}\over\pa\ph_x^\a}\bigg]_x+
[\,\dots\,]_{tt}+[\,\dots\,]_{xx}+[\,\dots\,]_{xt}=0,\qquad
\a=1,\dots,m\.\tag8
$$

The equations of motion have the form
$$
\frac{\pa\Cal L}{\pa a^q}-\frac\pa{\pa t}\,
\frac{\pa\Cal L}{\pa a_t^q}-\frac\pa{\pa x}\,
\frac{\pa\Cal L}{\pa a_x^q}+\dots=0,\qquad q=1,\dots,N.\tag9
$$

Assume that the Lagrangian is nondegenerate. Then we can proceed to
the Hamiltonian formalism by setting (the method of Ostrogradskii
applied to the case of fields)
$$
\alignat3
q_1^\a(x)&=\ph^\a(x),&\;&\dots,&\;q_{n_\a}^\a(x)&=\ph_{(n_\a-1)t}(x),
\tag10\\
r_1^i(x)&=u^i(x),&\;&\dots,&\;r_{\ovl11{\ssize n}_i}(x)&
=u_{(\ovl11{\ssize n}-1)t}^i(x)\.\tag11
\endalignat
$$
(Here $n_\a$ and $\ovl11n_i$ are the highest orders of time
derivatives of $\ph^\a$ and $u^i$, respectively, occurring in the
Lagrangian. Since we can perform linear transformations of the
coordinates $\ph^\a$ and $u^i$ (separately), it follows that in the
generic case all $n_\a$ (and all $\ovl11n_i$) are the same.) Let us
introduce the momenta
$$
\align
p_1^\a(x)&=\frac{\dt L}{\dt\ph_t^\a(x)}-
\dots+(-1)^{n_\a-1}\bigg[\frac{\dt L}{\dt\ph_{n_\a t}^\a(x)}
\bigg]_{(n_\a-1)t},\\
\dots\dots&\dots\dots\dots\dots\dots\dots\dots
\dots\dots\dots\dots\dots\dots\dots\dots\tag12\\
p_{n_\a}^\a(x)&=\frac{\dt L}{\dt\ph_{n_\a t}(x)},\\
s_1^i(x)&=\frac{\dt L}{\dt u_t^i(x)}-\dots+ (-1)^{\ovl11{\ssize n}_i-1}
\bigg[\frac{\dt L}{\dt u_{\ovl11{\ssize n}_it}^i(x)}
\bigg]_{(\ovl11{\ssize n}_i-1)t},\\
\dots\dots&\dots\dots\dots\dots\dots\dots\dots
\dots\dots\dots\dots\dots\dots\dots\tag13\\
s_{\ovl11{\ssize n}_i}^i(x)&=\frac{\dt L}{\dt u_{\ovl11{\ssize n}_it}^i(x)},
\endalign
$$
where
$$
\frac{\dt L}{\dt a_{nt}^q(x)}\equiv\frac{\pa\Cal L}{\pa a_{nt}^q}\(x)
-\frac\pa{\pa x}\,\frac{\pa\Cal L}{\pa a_{nt,x}^q}\(x)+\dots\.
$$

The equations of motion (9) in the variables $q_{\nu_\a}^\a(x)$,
$r_{\mu_i}^i(x)$, $p_{\nu_\a}^\a(x)$, $s_{\mu_i}^i(x)$ are
Hamiltonian. The Hamiltonian is equal to
$$
H=\int\bigg[\sum_{\a,\nu_\a}p_{\nu_\a}^\a(x\)
\dot q_{\nu_\a}^\a(x)+\sum_{i,\mu_i}s_{\mu_i}^i(x\)
\dot r_{\mu_i}^i(x)\bigg]dx-L,\tag14
$$
and the Poisson bracket has the form
$$
\{q_{\nu_\a}^\a(x),p_{\nu_\be}^\be(y)\}=\dt^{\a\be}\dt_{\nu_\a\nu_\be}
\dt(x-y),\quad\{r_{\mu_i}^i(x),s_{\mu_j}^j(y)\}=
\dt^{ij}\dt_{\mu_i\mu_j}\dt(x-y)\tag15
$$
(all other brackets are zero).

More generally, if the conservation law corresponding to the symmetry
of the Lagrangian with respect to a one-parameter transformation
family can be put in the form $P_t=Q_x$ (perhaps, ambiguously,
because of the presence of mixed $t,x$-derivatives, as is the case in
(6), (7), and (8)), then the functional $\int P\,dx$ (defined
unambiguously) generates this transformation family in the
Hamiltonian structure (15).

\rk{Remark} Very frequently, an originally degenerate Lagrangian
becomes nondegenerate after a suitable rotation in the plane $(x,t)$
(for example, this is the case for the Korteweg-de Vries (KdV) and
nonlinear Schr\"odinger (NLS) equations).
\endrk

Let us describe the set of functions to be considered in the sequel.
Let $\ovl11a(x)$ satisfy the conditions
$$
\exists\,k>0,\;\wh S\in T^m\quad\forall x\quad
\ovl11a(x+2\pi/k)=\wh S\ovl11a(x),\tag16
$$
that is, in the variables $(u^i,\ph^\a)$ we have
$$
u^i(x+2\pi/k)=u^i(x),\qquad\ph^\a(x+2\pi/k)=\ph^\a(x)+\mu^\a,\tag17
$$
and let the initial data be given in the form
$$
\forall x\quad u_{nt}^i(x+2\pi/k)=u_{nt}^i(x),\quad
\ph_{nt}^\a(x+2\pi/k)=\ph_{nt}^\a(x),\qquad n\ge1\.\tag18
$$

Since ${\Cal L}$ depends only on the derivatives of $\ph^\a$, the
evolution equations (9) preserve each of the families (17) and
property (18).

In terms of the Hamiltonian variables conditions (17) and (18)
acquire the form
$$
\gather
q_1^\a(x+2\pi/k)=q_1^\a(x)+\mu^\a,\quad
q_{\nu_\a}^\a(x+2\pi/k)=q_{\nu_\a}^\a(x),\qquad\nu_\a>1,\tag19\\
r_{\mu_i}^i(x+2\pi/k)=r_{\mu_i}^i(x),\quad
p_{\nu_\a}^\a(x+2\pi/k)=p_{\nu_\a}^\a(x),\quad
s_{\mu_i}^i(x+2\pi/k)=s_{\mu_i}^i(x)\.\tag20
\endgather
$$

On the family (19), (20) we seek extremals of all possible
functionals of the form
$$
\la_HH+\la_PP+\sum_\a\la_\a
I^\a,\qquad\la_H,\,\la_P,\,\la_\a=\op{const},\tag21
$$
where $H$ is the Hamiltonian, $P$ is the integral of momentum, and
$I^\a$ are the functionals that generate the shifts of the
corresponding angles $\ph^\a$. Thus, we consider the equation
$$
\dt\bigg[\la_HH+\la_PP+\sum_\a\la_\a I^\a\bigg]=0\.\tag22
$$

We assume that for any $k$, $\mu^\a$, $\la_H$, $\la_P$, and $\la_\a$
the solution to Eq.~(22) is unique modulo a translation along the
$x$-axis and the action of a transformation in $T^m$.

\rk{Comment} If for any $k>0$, $\mu^\a$, $\la_H$, $\la_P$, and
$\la_\a$ there exists a solution
$\ovl11a_{(k,\ovl11{\ssize\mu},\ovl11{\ssize\la})}(x)$,
then\linebreak
$\ovl11a_{(k/n,\ovl11{\ssize\mu}n,\ovl11{\ssize\la})}(x)$ is also a
solution for any positive integer $n$ and to each tuple
$(k,\ovl11{\ssize\mu},\ovl11{\ssize\la})$ there corresponds countably
many such solutions. Accordingly, we assume that our family of
solutions has just this very form, and to each
$\ovl11a_{(k,\ovl11{\ssize\mu},\ovl11{\ssize\la})}(x)$ we assign the
greatest admissible $k$ and the smallest admissible $\ovl11\mu$.
\endrk

The solutions to Eq.~(22) depend on the parameters $k$, $\mu^\a$,
$\la_P/\la_H$, $\la_\a/\la_H$, $\ph_0^\a$, and $\th_0$, where
$\ph_0^\a$ and $\th_0$ are the initial phases corresponding to the
action of $T^m$ and to the translations along the $x$-axis. The
averaging affects the $2m+2$ parameters $k$, $\mu^\a$, $\la_P/\la_H$,
and $\la_\a/\la_H$; the averaged variables do not depend on
$\ph_0^\a$ and $\th_0$, so that these last variables do not occur in
the equations of slow modulations.

Let us transform system (14), (15) into a new Hamiltonian system by
setting
$$
q^\a(x)=q_{1x}^\a(x),\quad p^\a(x)=p_1^\a(x)\tag23
$$
(this substitution is well-defined, since the densities of the
functionals $H$, $P$, and $I^\a$ depend only on the derivatives of
$q_1^\a$ with respect to $x$).

The new Poisson bracket has the form
$$
\gather
\{q_{\nu_\a}^\a(x),p_{\nu_\be}^\be(y)\}
=\dt^{\a\be}\dt_{\nu_\a\nu_\be}\dt(x-y),\qquad\nu_\a,\nu_\be\ge2,\tag24\\
\{r_{\mu_i}^i(x),s_{\mu_j}^j(y)\}=\dt^{ij}\dt_{\mu_i\mu_j}\dt(x-y),\quad
\{q^\a(x),p^\be(y)\}=\dt^{\a\be}\dt'(x-y)\tag25
\endgather
$$
(all other brackets are zero).

Thus, $\int q^\a(x)\,dx$ and $\int p^\be(x)\,dx$ are annihilators of
the bracket (24), (25), so that every Hamiltonian system with
translation-invariant Hamiltonian has $2m+2$ first integrals.

\tm{Lemma 1} The family of periodic traveling waves of the
Hamiltonian system\/ \rom{(14)}, \rom{(23)--(25)} can be obtained
from the family\/ \rom{(19)}, \rom{(20)}, \rom{(22)} by factorization
with respect to the initial phases $\ph_0^\a$.
\endtm

\pf{Proof} It is easy to see that conditions (19) and (20) are
equivalent to the $2\pi/k$-periodicity of the functions occurring in
(23)--(25).

It follows from conditions (22) that
$q_{1t}^\a=-(\la_P/\la_H\)q_{1x}^\a-\la_\a/\la_H$ and that
$\xi_t=-(\la_P/\la_H\)\xi_x$ for the other variables. By
differentiating the first equation with respect to~$x$ and by
substituting $q_{1x}^\a=q^\a(x)$ we obtain
$q_t^\a=-(\la_P/\la_H\)q_x^\a$; thus, after factorization with
respect to the initial phases $\ph_0^\a$ we obtain exactly the family
of periodic traveling waves of system (14), (23)--(25). The lemma is
proved.
\endpf

\tm{Corollary} The family of periodic traveling waves of system\/
\rom{(14)}, \rom{(23)--(25)} depends on $(2m+2)$ parameters (not
including the initial phase $\th_0$). The averaging of the
conservation laws for the energy, momentum, and $2m$ annihilators of
the bracket\/ \rom{(24)}, \rom{(25)}, yields the Whitham equations of
slow modulations.
\endtm

Since
$$
p_1^\a(x)=\frac{\pa\Cal L}{\pa\ph_t^\a}\(x),\qquad
\frac{\dt H}{\dt q^\a(x)}=\frac{\dt H}{\dt q_{1x}^\a(x)}
=-\frac{\pa\Cal L}{\pa\ph_x^\a}\(x)\tag26
$$
modulo total derivatives with respect to $x$ and $t$ and the energy
and momentum conservation laws have the form (6), (7), we see that
the equations of slow modulations have the form
$$
\multline
\bigg[\sum_{n\ge1}n\bigg(\sum_i\bigg\lan u_{nt}^i\,
\frac{\pa\Cal L}{\pa u_{nt}^i}\bigg\ran+
\sum_\a\bigg\lan\ph_{nt}^\a\,\frac{\pa\Cal L}{\pa\ph_{nt}^\a}
\bigg\ran\bigg)-\lan\Cal L\ran\bigg]_T\\
+\bigg[\sum_{n\ge1}n\bigg(\sum_i\bigg\lan u_{(n-1)x,t}^i\,
\frac{\pa\Cal L}{\pa u_{nx}^i}\bigg\ran+\sum_\a\bigg\lan
\ph_{(n-1)x,t}\,\frac{\pa\Cal L}{\pa\ph_{nx}^\a}\bigg\ran\bigg)\bigg]_X=0,
\endmultline\tag27
$$
\vskip-12pt
$$
\multline
\bigg[\sum_{n\ge1}n\bigg(\sum_i\bigg\lan u_{(n-1)t,x}^i
\frac{\pa\Cal L}{\pa u_{nt}^i}\bigg\ran+
\sum_\a\bigg\lan\ph_{(n-1)t,x}^\a\,
\frac{\pa\Cal L}{\pa\ph_{nt}^\a}\bigg\ran\bigg)\bigg]_T\\
+\bigg[\sum_{n\ge1}n\bigg(\sum_i\bigg\lan u_{nx}^i\,
\frac{\pa\Cal L}{\pa u_{nx}^i}\bigg\ran+\sum_\a\bigg\lan
\ph_{nx}^\a\,\frac{\pa\Cal L}{\pa\ph_{nx}^\a}\bigg\ran\bigg)
-\lan\Cal L\ran\bigg]_X=0,
\endmultline\tag28
$$
\vskip-14pt
$$
\gather
\bigg\lan\frac{\pa\Cal L}{\pa\ph_t^\a}
\bigg\ran_T+\bigg\lan\frac{\pa\Cal L}{\pa\ph_x^\a}\bigg\ran_X=0,\tag29\\
\lan q^\a\ran_T=\bigg\lan\frac{\dt H}{\dt p^\a(x)}\bigg\ran_X,\tag30
\endgather
$$
where $\lan\,\dots\,\ran$ denotes the averaging on the family (19),
(20), (22) (or, which is the same, on the family of periodic
traveling waves of system (14), (23)--(25)).

It is easy to see that Eqs.~(29) correspond to the conservation laws
(8).

Let us proceed to the averaging method for the Lagrangian.

Following Whitham, we seek solutions to Eqs.~(9) in the form
$$
u^i(x,t)=\Phi^i(\th),\quad\ph^\a(x,t)=\Psi^\a(\th)+\e^\a,\tag31
$$
where $\th=kx+\om t$ is the phase, $\e^\a=\be^\a x+\ga^\a t$ are
pseudophases, and $\Phi^i(\th)$ and $\Psi^\a(\th)$ are
$2\pi$-periodic functions.

Next, we set
$$
\Cal L(u^i,u_t^i,u_x^i,\dots,\ph_t^\a,\ph_x^\a,\ph_{tt}^\a,
\ph_{xx}^\a,\dots) =\Cal L(\Phi^i,\om\Phi_{\th}^i,k\Phi_{\th}^i,\dots,
\om\Psi_{\th}^\a+\ga^\a,k\Psi_{\th}^\a+\be^\a,\om^2\Psi_{\th\th}^\a,
k^2\Psi_{\th\th}^\a,\dots),
$$
where $\Phi^i=\Phi^i(\th,k,\om,\be^\a,\ga^\a)$,
$\Psi^\a=\Psi^\a(\th,k,\om,\be^\a,\ga^\a)$, construct the averaged
Lagrangian
$$
\ovL=\frac1{2\pi}\int_0^{2\pi}\Cal L(\th,k,\om,
\be^\a,\ga^\a)\,d\th=\ovL(k,\om,\be^\a,\ga^\a)\equiv\ovL(\th_X,
\th_T,\e_X^\a,\e_T^\a),\tag32
$$
and obtain the averaged equations
$$
[\ovL_{\th_X}]_X+[\ovL_{\th_T}]_T=0,\qquad
[\ovL_{\e_X^\a}]_X+[\ovL_{\e_T^\a}]_T+0,\tag33
$$
or, in the variables $k$, $\om$, $\be^\a$, $\ga^\a$,
$$
\alignat2
[\ovL_\om]_T+[\ovL_k]_X&=0,&\qquad k_T&=\om_X,\tag34\\
[\ovL_{\ga^\a}]_T+[\ovL_{\be^\a}]_X&=0,&\qquad\be_T^\a&=\ga_X^\a\.\tag35
\endalignat
$$

Since the Lagrangian $\ovL$ is independent of time and
translation-invariant, Eqs.~(34), (35) admit energy and momentum
conservation laws. On replacing Eqs.~(34) by these laws, we obtain
the averaged system in the form
$$
\gather
\bigg[\om\ovL+\sum_\a\ga^\a\ovL_{\ga^\a}-\ovL\bigg]_T+
\bigg[\om\ovL_k+\sum_\a\ga^\a\ovL_{\be^\a}\bigg]_X=0,\tag36\\
\bigg[k\ovL_\om+\sum_\a\be^\a\ovL_{\ga^\a}\bigg]_T+\bigg[k\ovL+
\sum_\a\be^\a\ovL_{\be^\a}-\ovL\bigg]_X=0,\tag37\\
[\ovL_{\ga^\a}]_T+[\ovL_{\be^\a}]_X=0,\tag38\\
\be_T^\a=\ga_X^\a\.\tag39
\endgather
$$

\tm{Lemma 2} The solution family\/ \rom{(31)} coincides with the
family\/ \rom{(19)}, \rom{(20)}, \rom{(22)}, and moreover,
$\be^\a=\lan q^\a\ran$, $\ga^\a=-(\la_P/\la_H\)\lan
q^\a\ran-\la_\a/\la_H$.
\endtm

\pf{Proof} The coincidence of these solution families is evident, and
(22) implies that
$$
\be^\a=(k/2\pi)[\ph^\a(x+2\pi/k)-
\ph^\a(x)]=\lan\ph_x^\a\ran=\lan q^\a\ran;
$$
similarly, we have $\ga^\a=\lan\ph_t^\a\ran=\lan q_{1t}^\a\ran$, but,
by virtue of (22), (22)
$q_{1t}^\a=-(\la_P/\la_H\)q_{1x}^\a-\la_\a/\la_H$, that is,
$$
\ga^\a=\lan-(\la_P/\la_H\)q_{1x}^\a-\la_\a/\la_H\ran
=-(\la_P/\la_H\)\lan q^\a\ran-\la_\a/\la_H\.
$$

The lemma is proved.
\endpf

It is also obvious that the averaging with respect to $\th$ is
equivalent to the averaging with respect to $x$ and $t$.

\tm{Theorem 1} Equations\/ \rom{(27)--(30)} coincide with
Eqs.\/~\rom{(36)--(39)}.
\endtm

\pf{Proof} Let us introduce the functions
$$
\wt{\Cal L}_i=\frac{\pa\Cal L}{\pa u^i},\;\;
\wh{\Cal L}_i^n=\frac{\pa\Cal L}{\pa u_{nx}^i},\;\;
\hhL_\a^n=\frac{\pa\Cal L}{\pa\ph_{nx}^\a},\;\;
\Lh_i^n=\frac{\pa\Cal L}{\pa u_{nt}^i},\;\;
\Lhh_\a^n=\frac{\pa\Cal L}{\pa\ph_{nt}^\a},\qquad n\ge1\.
$$
For all these functions we set
$$
\Cal L_*^*=\Cal L_*^*(\Phi^i,\om\Phi_\th^i,
k\Phi_\th^i,\dots,\om\Psi_\th^\a+\ga^\a,
k\Psi_\th^\a+\be^\a,\om^2\Psi_{\th\th}^\a,k^2\Psi_{\th\th}^\a,\dots),
$$
where $\Phi^i=\Phi^i(\th,k,\om,\ovl10\be,\ov\ga)$,
$\Psi^\a=\Psi^\a(\th,k,\om,\ovl10\be,\ov\ga)$, and define the action
$$
\wt S=\frac1{2\pi}\int_0^{2\pi}\Cal L(\Phi^i,\om
\Phi_\th^i,k\Phi_\th^i,\dots,\om\Psi_\th^\a+\ga^\a,
k\Psi_\th^\a+\be^\a,\om^2\Psi_{\th\th}^\a,
k^2\Psi_{\th\th}^\a,\dots)\,d\th\.
$$
(Obviously, $\wt S=\lan\Cal L\ran$.)

It is easy to verify that
$$
\frac{\dt\wt S}{\dt\Phi^i(\th)}=0,\qquad
\frac{\dt\wt S}{\dt\Psi^\a(\th)}=0\tag40
$$
on the solution family (31). Furthermore, it is obvious that
$$
\align
\ovL_\om&=\sum_{n\ge1}n\bigg(\sum_i\om^{n-1}
\lan\Phi_{n\th}^i\Lh_i^n\ran+\sum_\a\om^{n-1}\lan
\Psi_{n\th}^\a\Lhh_\a^n\ran\bigg)\\
&\qquad+\int_0^{2\pi}\sum_i\frac{\dt\wt S}{\dt\Phi^i(\th)}\,
\Phi_\om^i\,d\th+\int_0^{2\pi}\sum_\a
\frac{\dt\wt S}{\dt\Psi^\a(\th)}\,\Psi_\om^\a\,d\th,\\
\ovL_k&=\sum_{n\ge1}n\bigg(\sum_ik^{n-1}\lan
\Phi_{n\th}^i\wh{\Cal L}_i^n\ran+\sum_\a
k^{n-1}\lan\Psi_{n\th}^\a\hhL_\a^n\ran\bigg)\\
&\qquad+\int_0^{2\pi}\sum_i\frac{\dt\wt S}{\dt
\Phi^i(\th)}\,\Phi_k^i\,d\th+\int_0^{2\pi}\sum_\a
\frac{\dt\wt S}{\dt\Psi^\a(\th)}\,\Phi_k^\a\,d\th,\\
\ovL_{\be^\nu}&=\lan\hhL_\nu^1\ran
+\int_0^{2\pi}\sum_i\frac{\dt\wt S}{\dt\Phi^i(\th)}\,
\Phi_{\be^\nu}^i\,d\th+\int_0^{2\pi}\sum_\a
\frac{\dt\wt S}{\dt\Psi^\a(\th)}\,\Psi_{\be^\nu}^\a\,d\th,\\
\ovL_{\ga^\nu}&=\lan\Lhh_\nu^1\ran
+\int_0^{2\pi}\sum_i\frac{\dt\wt S}{\dt\Phi^i(\th)}\,
\Phi_{\ga^\nu}^i\,d\th+\int_0^{2\pi}\sum_
\a\frac{\dt\wt S}{\dt\Psi^\a(\th)}\,\Psi_{\ga^\nu}^\a d\th
\endalign
$$
(here notation such as $\ovL$ and $\lan\Cal L\ran$ is used for the
averaged variables).

\baselineskip=11.7pt

By (40), we have
$$
\gather
\ovL_\om=\sum_{n\ge1}n\bigg(\sum_i\om^{n-1}
\lan\Phi_{n\th}^i\Lh_i^n\ran+\sum_\a\om^{n-1}\lan\Psi_{n\th}^\a
\Lhh_\a^n\ran\bigg),\\
\ovL_k=\sum_{n\ge1}n\bigg(\sum_ik^{n-1}\lan
\Phi_{n\th}^i\wh{\Cal L}_i^n\ran
+\sum_\a k^{n-1}\lan\Psi_{n\th}^\a\hhL_\a^n\ran\bigg),\\
\ovL_{\be^\a}=\lan\hhL_\a^1\ran,\qquad\ovL_{\ga^\a}=\lan\Lhh_\a^1\ran\.
\endgather
$$
It follows from Eqs.~(27)--(30) and from Lemma 2 that
$$
\split
&\sum_{n\ge1}n\bigg(\sum_i\bigg\lan u_{nt}^i\,
\frac{\pa\Cal L}{\pa u_{nt}^i}\bigg\ran+\sum_\a
\bigg\lan\ph_{nt}^\a\,\frac{\pa\Cal L}{\pa\ph_{nt}^\a}\bigg\ran\bigg)
-\lan\Cal L\ran\\
&\qquad=\sum_{n\ge1}n\bigg(\sum_i\lan\om^n\Phi_{n\th}^i
\Lh_i^n\ran\bigg)+\sum_{n\ge1}n\bigg(\sum_\a\lan\om^n
\Psi_{n\th}^\a\Lhh\a^n\ran\bigg)+\sum_\a\lan\ga^\a
\Lhh_\a^1\ran-\lan\Cal L\ran\\
&\qquad=\om\ovL_\om+\sum_\a\ga^\a\ovL_{\ga^\a}-\ovL,\\
&\sum_{n\ge1}n\bigg(\sum_i\bigg\lan u_{(n-1)x,t}^i\,
\frac{\pa\Cal L}{\pa u_{nx}^i}\bigg\ran+\sum_\a\bigg\lan
\ph_{(n-1)x,t}^\a\,\frac{\pa\Cal L}{\pa\ph_{nx}^\a}\bigg\ran\bigg)\\
&\qquad=\sum_{n\ge1}n\bigg(\sum_i\lan k^{n-1}\om
\Phi_{n\th}^i\wh{\Cal L}_i^n\ran+\sum_\a
\lan k^{n-1}\om\Psi_{n\th}^\a\hhL_\a^n\ran\bigg)+
\sum_\a\lan\ga^\a\hhL_\a^1\ran\\
&\qquad=\om\ovL_k+\sum_\a\ga^\a\ovL_{\be^\a};
\endsplit
$$
thus, (27) coincides with (36), and
$$
\split
&\sum_{n\ge1}n\bigg(\sum_i\bigg\lan u_{(n-1)t,x}\,\frac{\pa\Cal L}
{\pa u_{nt}^i}\bigg\ran+\sum_\a\bigg\lan\ph_{(n-1)t,x}^\a\,
\frac{\pa\Cal L}{\pa\ph_{nt}^\a}\bigg\ran\bigg)\\
&\qquad=\sum_{n\ge1}n\bigg(\sum_i\lan\om^{n-1}k\Phi_{n\th}^i
\Lh_i^n\ran+\sum_\a\lan\om^{n-1}k\Psi_{n\th}^\a\Lhh_\a^n\ran\bigg)
+\sum_\a\lan\be^\a\Lhh_\a^1\ran\\
&\qquad=k\ovL_\om+\sum_\a\be^\a\ovL_{\ga^\a},\\
&\sum_{n\ge1}n\bigg(\sum_i\bigg\lan u_{nx}^i\,
\frac{\pa\Cal L}{\pa u_{nx}^i}\bigg\ran
+\sum_\a\bigg\lan\ph_{nx}^\a\,\frac{\pa\Cal L}{\pa\ph_{nx}^\a}\bigg\ran
\bigg)-\lan\Cal L\ran\\
&\qquad=\sum_{n\ge1}n\bigg(\sum_i\lan k^n\Phi_{n\th}^i
\wh{\Cal L}_i^n\ran\bigg)+\sum_{n\ge1}\bigg(\sum_\a\lan k^n
\Psi_{n\th}^\a\hhL_\a^n\ran\bigg)+\sum_\a\lan\be^\a\hhL_\a^1\ran
-\lan\Cal L\ran\\
&\qquad=k\ovL_k+\sum_\a\be^\a\ovL_{\be^\a}-\ovL,
\endsplit
$$
which means that (28) coincides with (37).

Furthermore,
$$
\bigg\lan\frac{\pa\Cal L}{\pa\ph_t^\a}\bigg\ran=
\lan\Lhh_\a^1\ran=\ovL_{\ga^\a},\qquad
\bigg\lan\frac{\pa\Cal L}{\pa\ph_x^\a}\bigg\ran=
\lan\hhL_\a^1\ran=\ovL_{\be^\a};
$$
that is, (29) coincides with (38).

Now, by (22),
$$
\bigg\lan\frac{\dt H}{\dt p_1^\a(x)}\bigg\ran
=\bigg\lan-\frac{\la_P}{\la_H}\,\{q_1^\a(x),P\}-
\sum_\be\frac{\la_\be}{\la_H}\,\{q_1^\a(x),I^\be\}\bigg\ran
=\bigg\lan-\frac{\la_P}{\la_H}\,q_{1x}^\a
-\frac{\la_\a}{\la_H}\bigg\ran=-\frac{\la_P}{\la_H}\,
\lan q^\a\ran-\frac{\la_\a}{\la_H},
$$
and (30) coincides with (39) by virtue of Lemma 2.

The theorem is proved.\endpf

\bh 2. Additional conservation laws\endbh Let us now give an
independent proof of the fact that the averaging of the conservation
laws corresponding to energy, momentum, and $I^\a$ on the family
(19), (20), (22) and the equations $\be_T^\a=\ga_X^\a$ (which
correspond to the averaging of the annihilators $\int q^\a(x)\,dx$ of
the Poisson bracket (24), (25)) imply the ``conservation of waves''
$k_T=\om_X$. Furthermore, we shall prove that if the Hamiltonian
system (14), (15) has additional conservation laws $S_t=R_x$
corresponding to some fluxes on the space of $a^q(x)$ which preserve
the Lagrangian and commute with the fluxes generated by the integral
of momentum and by $I^\a$ (in this case their densities and the
fluxes themselves are independent of $\ph^\a$ and, in the variables
used in (24), (25), correspond to conservation laws for system (14),
(24), (25)), then the averaging of these additional conservation laws
yields equations consistent with (27)--(30).

\tm{Theorem 2} The averaging of the conservation laws for the energy,
momentum, and $2m$ annihilators of the Hamiltonian system\/
\rom{(14)}, \rom{(24)}, \rom{(25)} on the family\/ \rom{(19)},
\rom{(20)}, \rom{(22)} implies the conservation of waves $k_T=\om_X$.
\endtm

\pf{Proof} In accordance with the preceding, we consider the equations
$$
\dt\bigg[\la_HH+\la_PP+\sum_\a\la_\a I^\a\bigg]=0
$$
on the phase space of functions that satisfy the conditions
$$
u^i(x+2\pi/k)=u^i(x),\qquad\ph^\a(x+2\pi/k)=\ph^\a(x)+2\pi\be^\a\!/k\.
$$
We assume that the variations $\dt u^i(x)$, $\dt\ph^\a(x)$ are
uniformly bounded, that is, $u^i(x)+\dt u^i(x)$,
$\ph^\a(x)+\dt\ph^\a(x)$ belong to the same family (15) as $u^i(x)$,
$\ph^\a(x)$. Thus,
$$
\frac\dt{\dt a^q(x)}\int\Cal P\,dx=\frac{\pa\Cal P}{\pa a^q}\(x)
-\frac\pa{\pa x}\,\frac{\pa\Cal P}{\pa a_x^q}\(x)+\dots\.
$$

Any functional that commutes with the Hamiltonian, with the momentum,
and with $I^\a$ leaves each of the families (19), (20) invariant.

Let $\ovU=(U^1,\dots,U^{2m+2})$ be a collection of parameters
(excluding the phases $\ph_0^\a$ and $\th_0$) on the solution family
(19), (20), (22). Let $\vec\xi$ be a tangent vector to the level
surface $k=\op{const}$, $\be^\a=\op{const}$ in the space with
coordinates $(U^1,\dots,U^{2m+2})$. Then the function variations
correponding to the translation along $\vec\xi$ are uniformly
bounded. Set
$$
H=\int\Cal P_H\,dx,\quad P=\int\Cal P_P\,dx,\quad I^\a=\int\Cal P_\a\,dx
$$
(the corresponding conservation laws have the form $(\Cal
P_H)_t=(J_H)_x$, $(\Cal P_P)_t=(J_P)_x$, $(\Cal P_\a)_t=(J_\a)_x$).
By (22), we have
$$
\la_H(\ovU\)\pa_{\vec\xi}\>\lan\Cal P_H\ran
+\la_P(\ovU\)\pa_{\vec\xi}\>\lan\Cal P_P\ran
+\sum_\a\la_\a(\ovU\)\pa_{\vec\xi}\>\lan\Cal P_\a\ran=0
$$
for any $\vec\xi$ such that $\pa_{\vec\xi}\>k=0$ and
$\pa_{\vec\xi}\>\be^\a=0$. This equation is equivalent to the relation
$$
\la_H(\ovU)\,d\>\lan\Cal P_H\ran+\la_P(\ovU)\,d\>\lan\Cal P_P\ran
+\sum_\a\la_\a(\ovU)\,d\>\lan\Cal P_\a\ran=\mu(\ovU)\,dk
+\sum_\a\mu_\a(\ovU)\,d\be^\a
$$
for some functions $\mu(\ovU)$ and $\mu_\a(\ovU)$.

As was already shown, under conditions (19) and (20) the solutions to
(22) have the form (31), that is,
$$
u^i(x,t)=\Phi^i(\th),\quad\ph^\a(x,t)=\Psi^
\a(\th)+\e^\a,\qquad\th=kx+\om t,\;\e^\a=\be^\a x+\ga^\a t,
$$
where $\Phi^i(\th)$, $\Psi^\a(\th)$ are $2\pi$-periodic functions.

In the original Lagrangian formalism let us make the rotation by an
angle $\chi$ in the plane $(x,t)$:
$$
\acute x=x\cos\chi-t\sin\chi,\qquad\acute t=x\sin\chi+t\cos\chi\.
$$
The passage to the Hamiltonian structure will now give a new Poisson
bracket, in which the translations along the ``old'' $t$-axis will be
generated by the functional $\int(\Cal
P_H\cos\chi-J_H\sin\chi)\,d\acute x$, the translations along the old
coordinate $x$ by the functional $\int(\Cal
P_P\cos\chi-J_P\sin\chi)\,d\acute x$, and the action of $T^m$ by the
functionals $\int(\Cal P_\a\cos\chi-J_\a\sin\chi)\,d\acute x$, where
all densities and fluxes of the conservation laws are expressed via
the new coordinates $a^q(\acute x)$. (Indeed, $\Cal P_t=J_x\iff(\Cal
P\cos\chi-J\sin\chi)_{\acute t}=(\Cal P\sin\chi+J\cos\chi)_{\acute
x}$.)

However, the solution family (31) remains the same, and moreover,
$\acute k=k\cos\chi-\om\sin\chi$, $\acute\be^\a=\be^\a\cos\chi-\ga^\a
\sin\chi$, $\acute\om=k\sin\chi+\om\cos\chi$,
$\acute\ga^\a=\be^\a\sin\chi+\ga^\a\cos\chi$.

By writing out the conditions
$$
\wh A\dt\bigg[\la_H\int\Cal P_H\,dx+\la_P
\int\Cal P_P\,dx+\sum_\a\la_\a\int\Cal P_\a\,dx\bigg]=0
$$
in the new variables for some function in the family (31), we obtain
$$
\multline
\wh A(\chi\)\dt\bigg[\la_H\int(\Cal P_H\cos\chi-
J_H\sin\chi)\,d\acute x+\la_P\int(\Cal P_P\cos\chi-
J_P\sin\chi)\,d\acute x\\
+\sum_\a\la_\a\int(\Cal P_\a\cos\chi-J_\a\sin\chi)\,d\acute x\bigg]=0,
\endmultline
$$
where $\wh A$ and $\wh A(\chi)$ are the old and the new Hamiltonian
operators, respectively.

Since $\wh A(\chi)$ is nondegenerate, we obtain
$$
\multline
\dt\bigg[\la_H\int(\Cal P_H\cos\chi-J_H\sin
\chi)\,d\acute x+\la_P\int(\Cal P_P\cos\chi-J_P\sin\chi)\,d\acute x\\
+\sum_\a\la_\a\int(\Cal P_\a\cos\chi-J_\a\sin\chi)\,d{\acute x}\bigg]=0\.
\endmultline
$$

Since the averaging on the family (31) in any direction in the plane
$(x,t)$ (except for $\th=\op{const}$, $\e^\a=\op{const}$) gives the
same result, we can conclude, as above, that
$$
\split
&\la_H(\ovU)(d\>\lan\Cal P_H\ran\cos\chi-d\>\lan J_H\ran\sin\chi)
+\la_P(\ovU) (d\>\lan\Cal P_P\ran\cos\chi-d\>\lan J_P\ran\sin\chi)\\
&\qquad\qquad+\sum_\a\la_\a(\ovU)(d\>\lan\Cal P_
\a\ran\cos\chi-d\>\lan J_\a\ran\sin\chi)\\
&\qquad=\mu(\chi,\ovU)\,d\acute k+\sum_\a\mu_\a(\chi,\ovU)\,d\acute\be^\a,
\endsplit
$$
that is,
$$
\split
&\la_H(\ovU)(d\>\lan\Cal P_H\ran\cos\chi-d\>\lan J_H\ran\sin\chi)
+\la_P(\ovU)(d\>\lan\Cal P_P\ran\cos\chi-d\>\lan J_P\ran\sin\chi)
\hskip40pt\\
&\qquad\qquad+\sum_\a\la_\a(\ovU)(d\>\lan\Cal P_\a
\ran\cos\chi-d\>\lan J_\a\ran\sin\chi)\\
&\qquad=\mu(\chi,\ovU)(dk\cos\chi-d\om\sin\chi)+
\topsmash{\sum_\a}\mu_\a(\chi,\ovU)(d\be^\a\cos\chi-d\ga^\a\sin\chi)\.
\endsplit\hskip-50pt\tag41
$$
For $\chi=0$ and $\chi=\pi/2$ we obtain, respectively,
$$
\align
\la_H(\ovU)\,d\>\lan\Cal P_H\ran+\la_P(\ovU)\,d\>\lan\Cal P_P\ran
+\sum_\a\la_\a(\ovU)\,d\>\lan\Cal P_\a\ran
&=\mu(0,\ovU)\,dk+\sum_\a\mu_\a(0,\ovU)\,d\be^\a,\tag42\\
\la_H(\ovU)\,d\>\lan\Cal P_H\ran+\la_P
(\ovU)\,d\>\lan J_P\ran+\sum_\a\la_\a(\ovU)\,d\>\lan J_\a\ran
&=\mu(\pi/2,\ovU)\,d\om+\sum_\a\mu_\a(\pi/2,\ovU)\,d\ga^\a.\tag43
\endalign
$$
Let us multiply (42) by $\cos\chi$ and (43) by $\sin\chi$ and
subtract the results from (41). We obtain
$$
\multline
[\mu(\chi,\ovU)-\mu(0,\ovU)]\cos\chi\,dk-
[\mu(\chi,\ovU)-\mu(\pi/2,\ovU)]\sin\chi\,d\om\\
+\sum_\a[\mu_\a(\chi,\ovU)-\mu(0,\ovU)]\cos\chi\,d\be^\a
-\sum_\a[\mu_\a(\chi,\ovU)-\mu_\a(\pi/2,\ovU)]\sin\chi\,d\ga^\a=0\.
\endmultline
$$
Assuming that the differentials $dk$, $d\om$, $d\be^\a$, and
$d\ga^\a$ are linearly independent (on the space $(\ovU)$), we obtain
$\mu(\chi,\ovU)\equiv\mu(0,\ovU)=\mu(\pi/2,\ovU)=\mu(\ovU)$,
$\mu_\a(\chi,\ovU)\equiv
\mu_\a(0,\ovU)=\mu_\a(\pi/2,\ovU)=\mu(\ovU)$. Consequently, on the
space $(\ovU)$ we have
$$
\align
\la_H(\ovU)\,d\>\lan\Cal P_H\ran+\la_P(\ovU)\,d\>\lan\Cal P_P\ran+
\sum_\a\la_\a(\ovU)\,d\>\lan\Cal P_\a\ran
&=\mu(\ovU)\,dk+\sum_\a\mu_\a(\ovU)\,d\be^\a,\\
\la_H(\ovU)\,d\>\lan J_H\ran+\la_P(\ovU)\,
d\>\lan J_P\ran+\sum_\a\la_\a(\ovU)\,d\>\lan J_\a\ran
&=\mu(\ovU)\,d\om+\sum_\a\mu_\a(\ovU)\,d\ga^\a;
\endalign
$$
these relations readily imply the statement of the theorem. The
theorem is proved.
\endpf

\tm{Theorem 3} Suppose that for any tuple $\la_H$, $\la_P$, $\la_\a$,
$k$, $\be^\a$ system\/ \rom{(22)} has a unique solution in the
class\/ \rom{(19)}, \rom{(20)} modulo translations along the $x$-axis
and the action of $I^\a$. Let the Hamiltonian system\/ \rom{(14)},
\rom{(15)} have an additional first integral $S_t=R_x$, corresponding
to some flux on the space of $a^q(x)$, and suppose that this flux
commutes with the fluxes generated by $H$, $P$, and $I^\a$ and
preserves the action. Then Eqs.~\rom{(27)--(30)} imply that $\lan
S\ran_T=\lan R\ran_X$.
\endtm

\pf{Proof} The flux generated by the functional $\int S\,dx$
preserves the solution family (19), (20), (22) and can at most
generate a linear dependence of the phases $\ph_0^\a$, $\th_0$ on
time on this family. Since the Poisson bracket (15) is nondegenerate,
it follows that
$$
\dt\bigg[\int S\,dx+\zeta(\ovU)\int\Cal P_P\,dx+\sum_\a\zeta_\a(\ovU)
\int\Cal P_\a\,dx\bigg]=0
$$
on the family (19), (20), (22) for some function $\zeta(\ovU)$ and
$\zeta_\a(\ovU)$. Literally repeating the argument in the proof of
Theorem 2, we see that
$$
\align
d\>\lan S\ran+\zeta(\ovU)\,d\>\lan\Cal P_P\ran
+\sum_\a\zeta_\a(\ovU)\,d\>\lan\Cal P_\a\ran
&=\eta(\ovU)\,dk+\sum_\a\eta_\a(\ovU)\,d\be^\a,\\
d\>\lan R\ran+\zeta(\ovU)\,d\>\lan J_P\ran+\sum_
\a\zeta_\a(\ovU)\,d\>\lan J_\a\ran
&=\eta(\ovU)\,d\om+\sum_\a\eta_\a(\ovU)\,d\ga^\a
\endalign
$$
on the manifold $(\ovU)$ for some functions $\eta(\ovU)$,
$\eta_\a(\ovU)$. The desired assertion now follows readily from
Theorem 2. The theorem is proved.
\endpf

\bh3. Hamiltonian formalism\endbh Equations (34), (35) are
Hamiltonian with respect to the Poisson bracket
$$
\{\be^\a(X),\ovL_{\ga^\be}(Y)\}=\dt^{\a\be}\dt'(X-Y),\qquad
\{k(X),\ovL_\om(Y)\}=\dt'(X-Y)\tag44
$$
(all other brackets are zero); the Hamiltonian is equal to
$$
H=\int\bigg(\om\ovL_\om+\sum_\a\ga^\a\ovL_{\ga^\a} -\ovL\bigg)dX,\tag45
$$
the integral of momentum
$$
P=\int\bigg(k\ovL_\om+\sum_\a\be^\a\ovL_{\ga^\a}\bigg)dX
$$
generates the translation along the $X$-axis, and the functionals
$\int\be^\a(X)\,dX$, $\int\ovL_{\ga^\a}(X)\,dX$, $\int k(X)\,dX$,
$\int\ovL_\om(X)\,dX$ are annihilators of the bracket (44) (see
\cite6).

Here we shall prove that the bracket (44) coincides with the averaged
Dubrovin--Novikov bracket \cite{2, 3} for Eqs.~(27)--(30).

Let us describe the construction of the Dubrovin--Novikov bracket.
Let the original evolution system
$$
a_t(x)=K(a,a_x,\dots)=\{a(x),H[a]\},\qquad a=(a^q),\tag46
$$
be Hamiltonian with respect to the local translation-invariant
field-theoretic Poisson bracket
$$
\{a^q(x),a^p(y)\}=\sum_{k=0}^MB_k^{qp}(a(x),a_x(x),
\dots\)\dt^{(k)}(x-y),\tag47
$$
and let the Hamiltonian $H[a]$ be a local field functional,
$$
H[a]=\int h(a(x),a_x(x),\dots)\,dx\.\tag48
$$
Suppose that system (46) has $N$ pairwise commuting local integrals
$$
I^i[a]=\int\Cal P^i(a(x),a_x(x),\dots)\,dx,\quad i=1,\dots,N,\qquad
\{I^i,I^j\}=0\.\tag49
$$
(In our case, $N=2m+2$ and the integrals have the form $\int
q^\a(x)\,dx$, $\int p^\a(x)\,dx$, $\int\Cal P_P\,dx$, and $\int\Cal
P_H\,dx$.) Consider the pairwise brackets of the densities of the
integrals (49)
$$
\multline
\{\Cal P^i(a(x),a_x(x),\dots),\Cal P^j(a(y),a_y(y),\dots)\}\\
=\sum_{k\ge0}A_k^{ij}(a(x),a_x(x),\dots\)\dt^{(k)}(x-y),\qquad
i,j=1,\dots,2m+2\.
\endmultline\tag50
$$
By (49), $A_0^{ij}=\pa_xQ^{ij}$. The averaged Dubrovin--Novikov
bracket has the form (we denote $\lan{\Cal P}^i\ran=U^i$)
$$
\split
\{U^i(X),U^j(Y)\}&=\lan A_1^{ij}\ran(\ovU(X))\dt'(X-Y)
+\frac{\pa\lan Q^{ij}\ran(\ovU)}{\pa U^k}\,U_X^k\dt(X-Y)\hskip15pt\\
&\equiv g^{ij}(\ovU(X)\)\dt'(X-Y)+b_k^{ij}(\ovU(X)\)U_X^k\dt(X-Y),
\endsplit\hskip-50pt\tag51
$$
where $\lan\,\dots\,\ran$ stands for the averaging on the family of
periodic traveling waves of the Hamiltonian system (46), determined
by the conditions
$$
\dt\bigg[k\int\Cal P_H\,dx-\om\int\Cal P_P\,dx+
\sum_\a\bigg(\mu_\a\int q^\a(x)\,dx
+\la_\a\int p_\a(x)\,dx\bigg)\bigg]=0,\tag52
$$
where $k$ is the wave number, $\om$ is the frequency, and $\mu_\a$
and $\la_\a$ are arbitrary constants. Just as in the proof of Theorem
2, it follows that
$$
k(\ovU)\,d\>\lan\Cal P_H\ran-\om(\ovU)\,d\>\lan\Cal P_P\ran
+\sum_\a(\mu_\a(\ovU\)d\>\lan q^\a\ran+\la_\a
(\ovU)\,d\>\lan p^\a\ran)=\mu(\ovU)\,dk\tag53
$$
for some function $\mu(\ovU)$.

The theory of the brackets (51) is closely related to Riemannian
geometry; in particular, their skew symmetry implies that
$$
g^{ij}=g^{ji},\qquad b_k^{ij}+b_k^{ji}=\pa g^{ij}\!/\pa U^k,\tag54
$$
and it follows from the Leibniz identity that under invertible smooth
changes $U^i\to\wt U^i(\ovU)$ of the field variables the functions
$g^{ij}$ behave as the contravariant components of a metric tensor,
and the functions $\Gamma_{jk}^i=-g_{js}b_k^{si}$ behave as the
components of the differential-geometric connection consistent
(by~(54)) with the metric.

If $g^{i,j}$ is nondegenerate, then the Jacobi identity for the
bracket (51) is equivalent to the symmetry of the connection
$\Gamma_{jk}^i$ and to the vanishing of the curvature tensor of the
metric, $R_{jkl}^i\equiv0$ (see \cite{2, 3}).

However, the survey \cite3 does not contain the proof of the Jacobi
identity for the bracket (51) obtained by averaging (see \cite4). In
other words, it is not proved that the connection $\Gamma_{jk}^i$ is
symmetric and that the curvature tensor of the metric $g^{ij}(\ovU)$
is zero. By proving that the Dubrovin--Novikov bracket coincides with
the bracket (44) rewritten in the variables $(\lan q^\a\ran,\lan
p^\a\ran,\lan\Cal P_P\ran,\lan\Cal P_H\ran)$ (as was shown in the
proof of Theorem 1, we have $\lan q^\a\ran=\be^\a$ and $\lan
p^\a\ran=\ovL_{\ga^\a}$), we at the same time prove that for the case
in question the Dubrovin--Novikov bracket satisfies the Jacobi
identity.

\tm{Theorem 4} The Dubrovin--Novikov bracket\/ \rom{(51)} coincides
with the bracket\/ \rom{(44)} rewritten in the variables $(\lan
q^\a\ran,\lan p^\a\ran,\lan\Cal P_P\ran,\lan\Cal P_H\ran)$.
\endtm

\pf{Proof} Let the bracket (44) in the variables $(\lan q^\a\ran,\lan
p^\a\ran,\lan\Cal P_P\ran,\lan\Cal P_H\ran)$ have the form
$$
\{U^i(X),U^j(Y)\}=\tilde g^{ij}(\ovU(X)\)\dt'
(X-Y)+\tilde b_k^{ij}(\ovU(X)\)U_X^k\dt(X-Y)\.\tag55
$$
It is easy to verify \cite3 that the Dubrovin--Novikov bracket can be
obtained as follows. With the bracket (47) we associate a bracket on
the space of fields $a^q(x,X)$ by setting
$$
\{a^q(x,X),a^p(y,Y)\}_0=\botsmash{\sum_{k=0}^M}
B_k^{qp}(a(x,X),a_x(x,X),\dots\)\dt^{(k)}(x-y\)\dt(X-Y),\tag56
$$
and then in the densities of the integrals $I_i(X)=\int\Cal
P_i(a(x,X),a_x(x,X),\dots)\,dx$ and in the bracket (56) we replace
the operator $\pa_x$ by $\pa_x+\e\pa_X$, where $\e\ll1$ is the ratio
of the ``rapid'' scale to the ``slow'' scale (for example, the ratio
of the wavelength to the typical length of variation of the
parameters $\ovU$). Then we evaluate the Poisson bracket of the
resultant integrals $I_\e^i(X)$ in the new Hamiltonian structure:
$$
\multline
\{a^q(x,X),a^p(y,Y)\}_\e\\
=\sum_{k=0}^MB_k^{qp}(a(x,X),a_x(x,X)+\e a_X(x,X),\dots)
\bigg(\frac\pa{\pa x}+\e\,\frac\pa{\pa X}\bigg)^k\dt(x-y\)\dt(X-Y)\.
\endmultline\tag57
$$
Set
$$
h_\e^{ij}(X,Y)=\{I_\e^i(X),I_\e^j(Y)\}_\e=\e h_1^{ij}(X,Y)+O(\e^2)
$$
and consider the restriction of the function
$h_1^{ij}(X,Y)=h_1^{ij}([a],X,Y)$ to the submanifold of functions
$a(x,X)$ such that for any fixed $X$ these functions, regarded as a
functions of $x$, belong to the family of traveling wave solutions to
system (46). The function $\lan{\Cal P}^i\ran(X)$, $\th_0(X)$ can
serve as coordinates on this manifold, and the Dubrovin--Novikov
bracket has the form
$$
\{\lan\Cal P^i\ran(X),\lan\Cal P^j(Y)\}=\tilde h_1^{ij}(X,Y),
$$
where $\tilde h_1^{ij}(X,Y)$ are the functions $h_1^{ij}(X,Y)$
restricted to the cited submanifold (it is easy to see that they are
independent of $\th_0(X)$). It readily follows that in this
Hamiltonian structure the functionals
$$
\int\lan\Cal P_H\ran\,dX,\quad\int\lan\Cal P_P\ran\,dX,\quad
\int\lan q^\a\ran\,dX,\quad\int\lan p^\a\ran\,dX\tag58
$$
generate the averaged equations (27)--(30), the translation along the
$X$-axis, and the ``zero'' fluxes, respectively on the space of
fields $U^i(X)$.

As was shown in the proof of Theorem 1, $\lan\Cal P_H\ran$, $\lan\Cal
P_P\ran$, and $\lan q^\a\ran$, $\lan p^\a\ran$ coincide,
respectively, with the densities of the Hamiltonian, of the momentum,
and of the annihilators of the Hamiltonian system (44), (45), and the
fluxes generated by the functionals (58) by virtue of the
Dubrovin--Novikov bracket coincide with the corresponding fluxes for
the bracket (44). Since the related equations in the coordinates
$\lan\Cal P^i\ran(X)=U^I(X)$ have the form $U^i_{t_j}=b_k^
{ij}(\ovU\)U_X^k$ and $U^i_{t_j}=\tilde b_k^{ij}(\ovU\)U_X^k$,
respectively, it readily follows that $b_k^{ij}=\tilde b_k^{ij}$ and
consequently, by~(54), $g^{ij}=\tilde g^{ij}+f^{ij}$, where
$f^{ij}=\op{const}$.

Furthermore, let
$$
\align
I_\e^i(X)&=I_0^i(X)+\e I_1^i(X)+O(\e^2),\\
\{\>\dots,\dots\}_\e&=\{\>\dots,\dots\}_0+\e\{\>\dots,\dots\}_1+O(\e^2);
\endalign
$$
then
$$
h_1^{ij}(X,Y)=\{I_0^i(X),I_1^j(Y)\}_0+
\{I_1^i(X),I_0^j(Y)\}_0+\{I_0^i(X),I_0^j(Y)\}_1,
$$
and, by (52) (we set $P=I^{2m+1}$, $H=I^{2m+2}$) on the cited
submanifold we have
$$
\sum_{i,j}\tau_i(X\)\tau_j(Y\)\tilde h_1^{ij}(X,Y)\equiv0,
$$
where $(\tau_i)=(\mu_1,\dots,\mu_m,\la_1,\dots,\la_m,-\om,k)$. Since
$\tilde h_1^{ij}(X,Y)$ has the form (51), we see that this relation
is equivalent to the relation
$$
\sum_{i,j}\tau_i(\ovU\)\tau_j(\ovU\)g^{ij}(\ovU)\equiv0;
$$
thus, $(\tau_i(\ovU))=(\mu_1(\ovU),\dots,\mu_m(\ovU),
\la_1(\ovU),\dots,\la_m(\ovU),-\om(\ovU),k(\ovU))$ is an isotropic
covector with respect to the metric $g^{ij}(\ovU)$. In view of (53),
we see that in the variables $(\be^\a,\ovL_{\ga^\a},k,\ovL_\om)$ this
covector has the coordinates $(0,\dots,0,0,\dots,0,\mu,0)$ and is
isotropic with respect to the metric (44). It follows that in the
variables $(U^i)$ the covector $(\mu_1(\ovU),\dots,\mu_m(\ovU),\la_
1(\ovU),\dots,\la_m(\ovU),-\om(\ovU),k(\ovU))$ is isotropic with
respect to both the metric $g^{ij}(\ovU)$ and the metric $\tilde
g^{ij}(\ovU)$, that is, it is isotropic with respect to $f^{ij}$.
However, $f^{ij}=\op{const}$, and $\mu_\a$, $\la_\a$, $\om$, $k$ can
be arbitrary by assumption. Thus, any covector is isotropic with
respect to $f^{ij}$; that is, $f^{ij}\equiv0$. The theorem is proved.
\endpf

\ex{Example} The multicomponent NLS-type equation
$$
iu_t^i+u_{xx}^i+V'(h\)u^i=0,\qquad h=\sum_i|u^i|^2,\;i=1,\dots,n,
$$
after the substitution $u^i=\sqrt{w^i}\exp\(i\int v^idx)$ takes the
form
$$
\gather
v_t^i=\bigg[\pa_x\bigg(\frac{w_x^i}{2w^i}\bigg)
+\bigg(\frac{w_x^i}{2w^i}\bigg)^2-(v^i)^2+
V'\bigg(\sum_iw^i\bigg)\bigg]_x,\tag59\\
w_t^i=-2\(w^iv^i)_x\.\tag60
\endgather
$$

The system has the Hamiltonian structure
$$
\gather
\{v^i(x),w^k(y)\}=\dt^{ik}\dt'(x-y),\\
H=\int\bigg[\sum_i\bigg(-\frac{(w_x^i)^2}{4w^i}-w^i(v^i)^2\bigg)
+V\bigg(\sum_iw^i\bigg)\bigg]dx,
\endgather
$$
and the integral of momentum has the form $P=\int\sum_iv^iw^i dx$.

The bracket has the form described in the preceding, and we can
proceed to a nondegenerate Lagrangian formalism in the system after
the substitution $w^i=q_x^i$ with allowance for the fact that
$v^i=-q_t^i/(2q_x^i)$ (thus, in accordance with the general
procedure, we regard $v^i$ as generalized momenta and express them
via $q_t^i$). The system takes the Lagrangian form with the Lagrangian
$$
\Cal L=\sum_i\bigg[-\frac{(q_t^i)^2}{4q_x^i}+
\frac{(q_{xx}^i)^2}{4q_x^i}\bigg]+V\bigg(\sum_iq_x^i\bigg).
$$

We find the family of traveling waves of system (59), (60) from the
conditions
$$
cv_x^i=\bigg[\frac{w_x^i}{2w^i}\bigg]_{xx}+
\bigg[\bigg(\frac{w_x^i}{2w^i}^2\bigg)\bigg]_x-
[(v^i)^2]_x+\bigg[V'\bigg(\sum_kw^k\bigg)\bigg]_x,\qquad
cw_x^i=-2(w^iv^i)_x\.
$$
It follows that
$$
cv^i=\bigg[\frac{w_x^i}{2w^i}\bigg]_x+
\bigg(\frac{w_x^i}{2w^i}\bigg)^2-(v^i)^2+
V'\bigg(\sum_kw^k\bigg)+A^i,\qquad cw^i+2w^iv^i=B^i,
$$
that is, $v^i=(B^i-cw^i)/(2w^i)$ and
$$
\frac{c^2}4+\bigg[\frac{w_x^i}{2w^i}\bigg]_x+
\bigg(\frac{w_x^i}{2w^i}\bigg)^2-\bigg(\frac{B^i}{2w^i}\bigg)^2
+V'\bigg(\sum_kw^k\bigg)+A^i=0\.\tag61
$$
System (61) has the first integral
$$
\sum_i\bigg(\frac{c^2}4\,w^i+\frac{(w_x^i)^2}{4w^i}
+\frac{(B^i)^2}{4w^i}+A^iw^i\bigg)+V\bigg(\sum_iw^i\bigg)=E
$$
and is Hamiltonian with generalized momenta $p^i= w_x^i/(2w^i)$,
Hamiltonian
$$
H=\sum_i\bigg(\frac{c^2}4\,w^i+w^i(p^i)^2+
\frac{(B^i)^2}{4w^i}+A^iw^i\bigg)+V\bigg(\sum_iw^i\bigg)
$$
and bracket $\{w^i,p^j\}=\dt^{ij}$.

By applying the canonical transformation $\ovl1{0.5}q^i=2\sqrt{w^i}$,
$\ovl10p^i=\sqrt{w^i}\>p^i$\<, we obtain
$$
H=\sum_i\bigg((\ovl10p^i)^2+\bigg(\frac{c\ovl1{0.5}q^i}4\bigg)^2
+\bigg(\frac{B^i}{\ovl1{0.5}q^i}\bigg)^2+
\frac{A^i(\ovl1{0.5}q^i)^2}4\bigg)
+V\bigg(\sum_i\frac{(\ovl1{0.5}q^i)^2}4\bigg),\quad
\{\ovl1{0.5}q^i,\ovl10p^j\}=\dt^{ij}.
$$
Thus, the periodic traveling waves of system (59), (60) correspond to
closed trajectories of a particle that moves in the $n$-dimensional
space with the potential
$$
\Pi(\ovl1{0.5}q)=\sum_i\bigg(\bigg(\frac{c\ovl1{0.5}q^i}4\bigg)^2
+\bigg(\frac{B^i}{\ovl1{0.5}q^i}\bigg)^2+\bigg(\frac{A^i\ovl1{0.5}q^i}
2\bigg)^2\bigg)+V\bigg(\sum_i\frac{(\ovl1{0.5}q^i)^2}4\bigg)
$$
with all possible $c$, $A^i$, $B^i$. Assuming that (as usual) closed
trajectories are isolated on each energy level, we obtain a
$(2n+2)$-parameter (excluding the initial phase shift $\th_0$) family
of traveling waves (or several such families) with the parameters
$c$, $A^i$, $B^i$, $E$, on which we can average the $2n+2$ first
integrals $P$, $H$, $\int v^idx$, and $\int w^idx$; all statements
proved above are valid for this family.\endex

In closing, the authors express their gratitude to S.~P.~Novikov for
his attention to the work, and also to I.~M.~Krichever,
E.~V.~Ferapontov, and O.~I.~Mokhov for useful remarks and fruitful
discussion in the course of the present study.

\Refs

\item{1.} G. Whitham, Linear and Nonlinear Waves, Wiley, New York
(1974).

\item{2.} B. A. Dubrovin and S. P. Novikov, ``Hamiltonian formalism
of one-dimensional systems of hydrodynamic type and the
Bogolyubov--Whitham averaging method,'' Dokl. Akad. Nauk SSSR,
{\bf270}, No.~4, 781--785 (1983).

\item{3.} B. A. Dubrovin and S. P. Novikov, ``Hydrodynamics of weakly
deformed soliton lattices. Differential geometry and Hamiltonian
theory,'' Uspekhi Mat. Nauk, {\bf44}, No.~6 (270), 29--98 (1989).

\item{4.} S. P. Novikov and A. Ya. Maltsev, ``The Liouville form of
averaged Poisson brackets,'' Uspekhi Mat. Nauk, {\bf48}, No.~1 (289),
155--156 (1993).

\item{5.} M. V. Pavlov, ``The Hamiltonian structure of Whitham's
equations,'' Uspekhi Mat. Nauk, {\bf49}, No.~1 (295), 219--220 (1994).

\item{6.} M. V. Pavlov, ``Double Lagrangian representation of the KdV
equation,'' Dokl. Ross. Akad. Nauk, {\bf339}, No.~2, 157--161 (1994).

\item{7.} S. P. Tsarev, ``On Poisson brackets and one-dimensional
Hamiltonian systems of hydrodynamic type,'' Dokl. Akad. Nauk SSSR,
{\bf282}, No.~3, 534--537 (1985).

\item{8.} I. M. Krichever, ``Averaging method for two-dimensional
`integrable' equations,'' Funkts. Anal. Prilo\-zhen., {\bf22}, No.~3,
37--52 (1988).

\item{9.} I. M. Krichever, ``Spectral theory of two-dimensional
operators and its applications,'' Uspekhi Mat. Nauk, {\bf44}, No.~2,
121--184 (1989).

\item{10.} M. J. Ablowitz and D. J. Benney, ``The evolution of
multiphase modes for nonlinear dispersive waves,'' Stud. Appl. Math.,
{\bf49}, No.~3, 225--238 (1970).

\item{11.} W. D. Hayes, ``Group velocity and nonlinear dispersive
wave propagation,'' Proc. Roy. Soc. London, {\bf332}, 199--221 (1973).

\item{12.} H. Flaschka, M. G. Forest, and D. W. McLaughlin,
``Multiphase averaging and the inverse spectral solution of the
Korteweg--de Vries equation,'' Comm. Pure Appl. Math., {\bf33},
No.~6, 739--784 (1980).

\item{13.} J. C. Luke, ``A perturbation method for nonlinear
dispersive wave problems,'' Proc. Roy. Soc. London Ser. A, {\bf292},
No.~1430, 403--412 (1966).

\item{14.} S. Yu. Dobrokhotov and V. P. Maslov, Finite-Gap Almost
Periodic Solutions in the WKB Approximation. Contemporary Problems in
Mathematics [in Russian], Vol.~15, Itogi Nauki i Tekhniki, VINITI,
Moscow (1980).

\endRefs

\transl{V. E. Nazaikinskii}
\enddocument